# Structural and optical properties of post annealed Mg doped ZnO thin films deposited by sol-gel method

Joydip Sengupta[a*], Arifeen Ahmed[a†] and Rini Labar[a†]

[a] Department of Physics, Sikkim Manipal Institute of Technology, Sikkim - 737136, India.

Thin films of magnesium doped zinc oxide (MZO) were synthesized using sol-gel method and annealed at different temperatures under ambient condition. The morphological properties of the post annealed MZO films were investigated using X-ray diffraction (XRD) while atomic force microscopy (AFM) was employed to probe the topographical alteration of the films. The optical properties of the post annealed MZO films were examined by UV-Visible spectroscopy and the Tauc method was used to estimate the optical band gap. The studies revealed that, with the rise of annealing temperature the crystallinity and the surface roughness of the MZO films were increased whereas the optical transmittance and the energy bandgap were decreased.

## 1. Introduction

A wide direct band gap (3.37 eV) and large exciton binding energy (60 meV) at room temperature have established ZnO as a promising material for optoelectronic devices such as fast ultraviolet detector, light emitting diode, laser diode etc. [1]. Nevertheless, there are few limitations in the application of pure ZnO for the integrated optical devices as they often require a much wider band gap [2]. Hence, the band gap tuning of ZnO becomes significant. The band gap of ZnO can be tailored by alloying ZnO with group II elements e.g. Be, Mg, Ca, Cd, and Sr [3]. Moreover, the radius of the **$Mg^{2+}$** ion (0.57 Å) closely matches with the radius of **$Zn^{2+}$** ion (0.60 Å), so the incorporation of **$Mg^{2+}$** ion into ZnO lattice is quite feasible, thus forming MZO. The enhanced optical band gap of MZO has been successfully employed to sense the mid and deep UV light [4]. The MZO thin films can be synthesized using various techniques such as RF sputtering, molecular-beam epitaxy, sol-gel, pulse laser deposition and chemical vapor deposition [2, 3, 4]. However, the sol-gel route offers many advantages like simple, inexpensive preparation of a large-area homogeneous thin film along with excellent compositional control, lower crystallization temperature and uniform film thickness. Furthermore, at different steps of fabrication, optoelectronic devices often require thermal treatments. So the study of the effect of thermal treatment on the structural and optical properties of MZO films requires serious attention for optoelectronic applications of these films.

---

[*] Corresponding author: E-mail: joydipdhruba@gmail.com  FAX: +91-33-2337-4637

[†] The authors A. Ahmed and R. Labar had contributed equally to this work

**2. Experimental procedure**

Zinc acetate dihydrate, Methanol and Diethanolamine (DEA) were used as starting material, solvent and stabilizer, respectively. First, **6.585 gm of** Zinc acetate dihydrate was dissolved in **40 ml** Methanol and then DEA was slowly added under magnetic stirring to prepare a solution of 0.75 M. Magnesium doping **of ZnO** was performed by adding Magnesium acetate tetrahydrate to **the Methanol along with the Zinc acetate dihydrate** in two different Mg/Zn ratios. For the sample MZO1, Mg/Zn ratio was 1.0 at.% and Mg/Zn ratio was 4.0 at.% for the sample MZO4. **For Mg doping of 1.0 at.% with respect to Zn, 0.0617 gm of Magnesium acetate tetrahydrate and for 4.0 at.% doping, 0.2468 gm of Magnesium acetate tetrahydrate was added to the solution.** The resulting mixture was stirred for 1 h at 65 °C, and then 3 h at room temperature to yield a clear and homogeneous solution. Afterwards the solution was **kept in a properly covered beaker for 48 h at room temperature for aging**, which served as the coating solution. **The quartz substrates were cleaned ultrasonically in DI water, Isopropyl alcohol and Acetone, successively**. Thin films of MZO were prepared by spin coating of the respective coating solutions onto pre-cleaned quartz substrates at rotation speed of 2000 rpm for 120 s in ambient condition. Later on the films were dried at 300 °C for 10 min in air to evaporate the solvent and organic residues. Finally these films were inserted into a resistive furnace and annealed in air at 400, 550 and 700 °C for 1 h. An AFM (Nanonics Multiview 4000[TM]) in intermittent contact mode, Bruker D8 diffractometer with Cu source and UV-Visible spectrophotometer (UV-1800, Shimadzu) were used to characterize the post annealed MZO films.

**3. Results and discussion**

The XRD spectra (Fig. 1) revealed the influence of annealing treatment in the temperature range of 400 - 700 °C on the structure of the MZO thin films. The X-ray pattern (Fig. 1) depicted that all the films were polycrystalline in nature with hexagonal wurtzite structure. Furthermore, for all the MZO films, neither MgO nor Mg phase was detected in the X-ray pattern. It was observed (Table 1) that the XRD peak related to (002) plane shifted systematically towards higher angle with the increase in annealing temperature i.e., from 34.47° to 34.57° for MZO1 and from 34.49° to 34.60° for MZO4. The reason behind the peak shifting was the change of stress, firstly due to the increase in annealing temperature [5, 6] additionally due to the increase in Mg concentration [7].

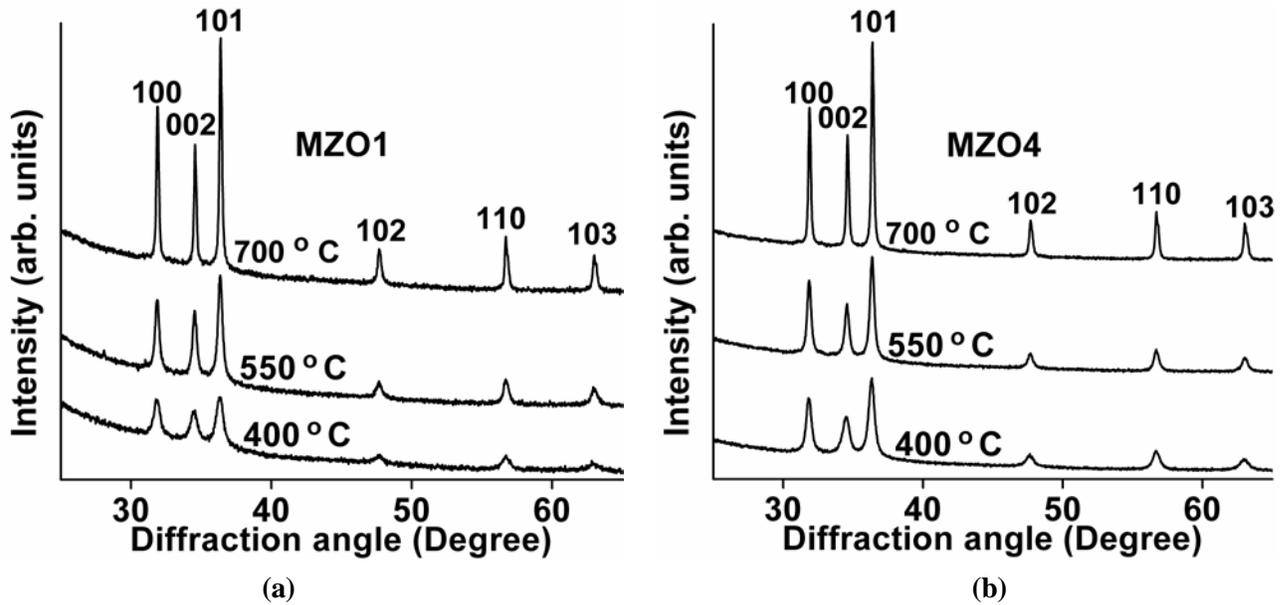

**Fig. 1.** X-ray diffraction spectra of MZO thin films deposited on quartz substrate after annealing at different temperatures: (a) MZO1 and (b) MZO4.

The **lower boundary of** the grain size ($D$) of the films was estimated using the full width at half maximum (FWHM) of (002) peak from the Scherrer's equation

$$D = \frac{K\lambda}{\beta \, Cos\theta} \qquad \ldots(1)$$

where $K = 0.9$ is the shape factor, $\lambda$ is the wavelength of incident X-ray, $\beta$ is the FWHM measured in radians and $\theta$ is the Bragg angle of diffraction peak. The XRD analysis (Table 1) revealed that, the FWHM of the peak corresponds to (002) plane was narrowed with increasing annealing temperature, indicating improvement in crystallinity [8]. It was also noticed (Table 1) that **lower boundary of the** grain size was gradually enlarged with enhancing annealing temperature for MZO films. This could be attributed to the coalescences of grains at higher annealing temperatures [8]. **Furthermore, estimation of the lower boundary of the grain size using the Scherrer's equation on the (002) peak implicates that, the results is also affected by the shift in the peak position due to stress in the film.** However, by comparing the **lower boundary of the** grain size of MZO1 and MZO4 for identical annealing temperature, it could be stated that Mg incorporation slightly deteriorates the crystallinity of the thin films [9].

**Table 1.** The data evaluated form the XRD, AFM and UV-Vis measurements of sol-gel derived MZO thin films after annealing at different temperatures

| Sample | Annealing temperature (°C) | FWHM of (002) peak (degree) | Position of (002) peak (2θ) | **Lower boundary of** grain size (nm) | RMS roughness (nm) | Optical bandgap (eV) |
|---|---|---|---|---|---|---|
| MZO1 | 400 | 0.5642 | 34.47 | 14.8 | 3.99 | 3.31 |
|  | 550 | 0.3715 | 34.53 | 22.4 | 8.39 | 3.28 |
|  | 700 | 0.2169 | 34.57 | 38.4 | 13.28 | 3.24 |
| MZO4 | 400 | 0.5750 | 34.49 | 14.5 | 6.08 | 3.40 |
|  | 550 | 0.3855 | 34.56 | 21.6 | 10.16 | 3.35 |
|  | 700 | 0.2287 | 34.60 | 36.4 | 18.94 | 3.26 |

Figure 2 shows the AFM micrographs of the surface profile for the post annealed MZO thin films on quartz substrate. AFM studies exhibited that the surface of all the films were highly dense and without any cracks. On close inspection of the AFM images, it was noticed that with the increase of annealing temperature the grain size became larger, which was consistent with the results of XRD. The topographical analysis revealed that (Table 1) the root mean square (RMS) roughness of the MZO thin films was monotonically increased with the increasing annealing temperature. This was attributed to the major grain growth which resulted in an increase of surface roughness [10].

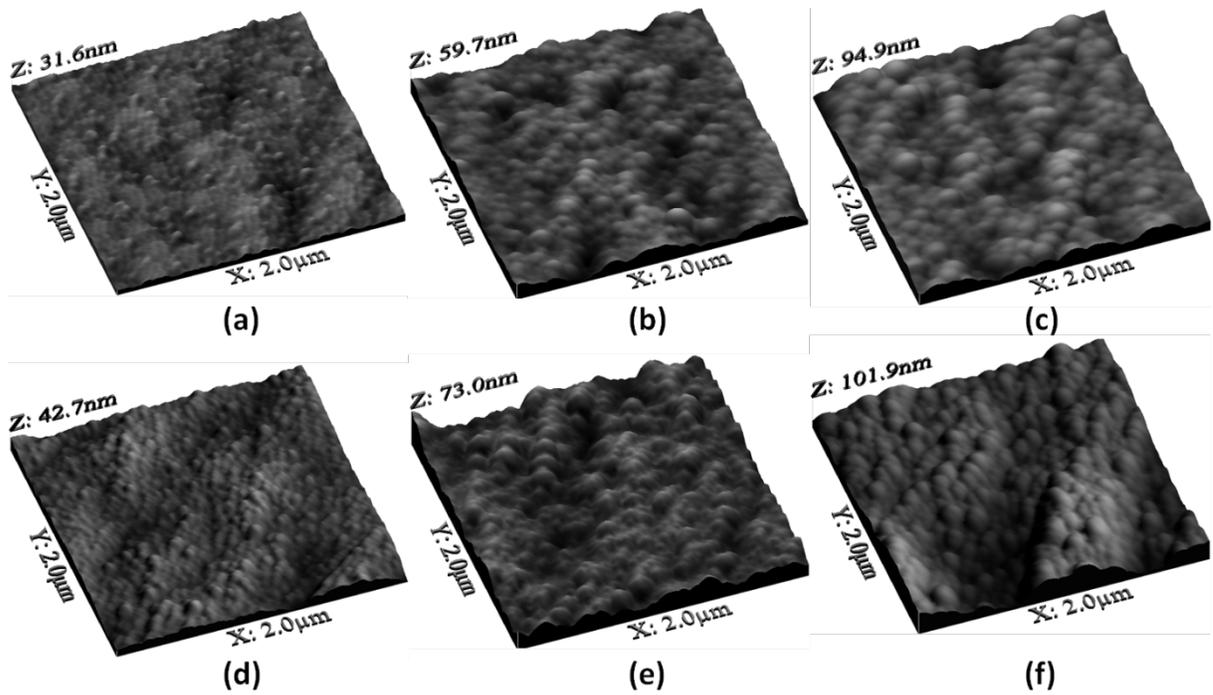

**Fig. 2.** Three dimensional AFM images of MZO thin films after annealing at different temperatures in air for 1 h (a) 400 °C, (b) 550 °C, (c) 700 °C for MZO1 film: (d) 400 °C, (e) 550 °C, (f) 700 °C for MZO4 film.

Figure 3a and b show the UV-Vis transmission spectra of the MZO thin films annealed at different temperatures with varying Mg concentrations. In this study, the transmittance of the samples was measured as a function of wavelength in the range of 200 to 800 nm in ambient condition. Spectrophotometric measurements of annealed MZO thin films (Fig. 3a and b) revealed that the transmittance was diminished with the rise in annealing temperature. This could be associated to the increased surface roughness upon annealing, which scattered the incident light [11].

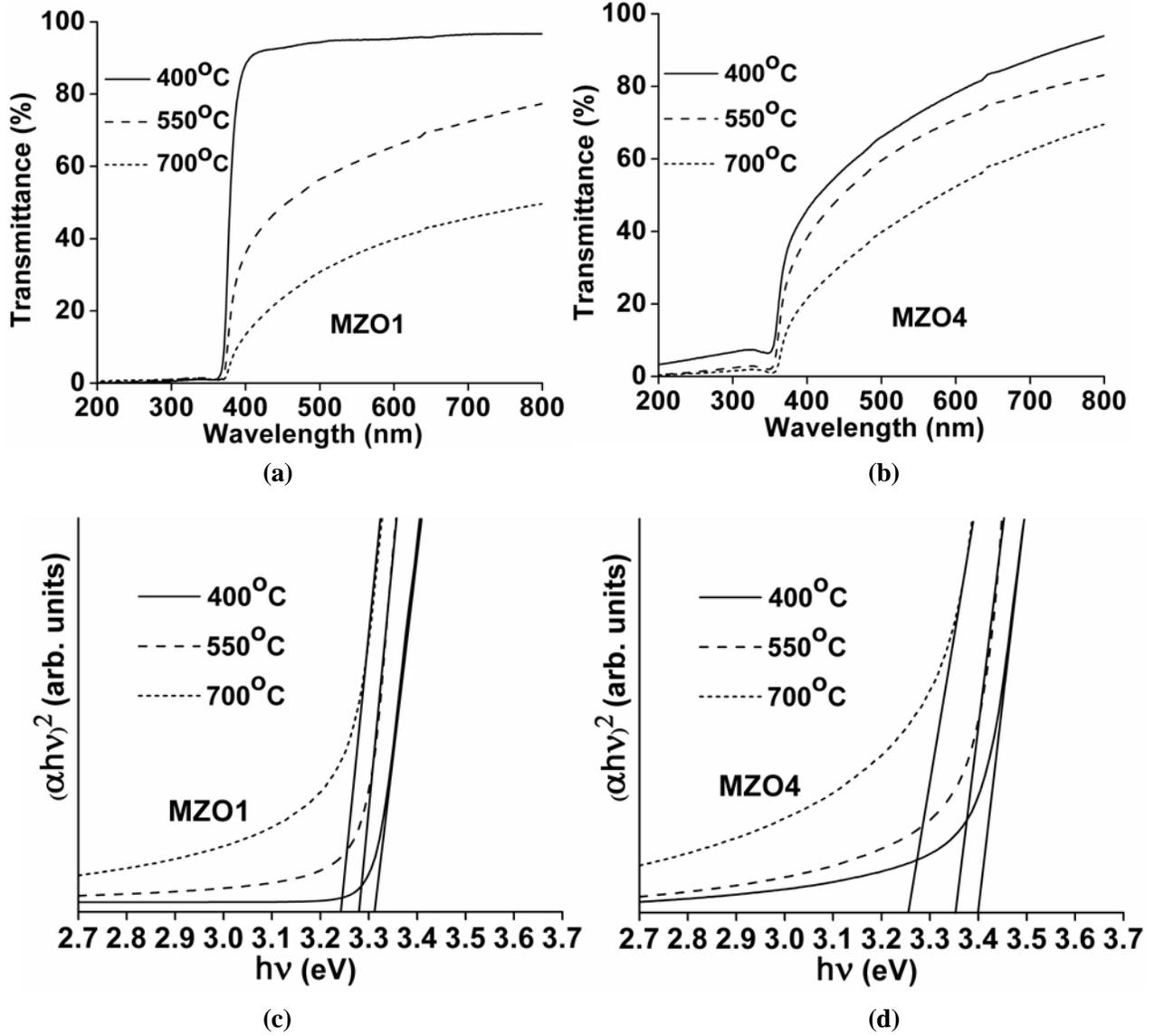

**Fig. 3.** Optical transmittance spectra of sol-gel derived MZO thin films after annealing at different temperatures. (a) MZO1 and (b) MZO4. Tauc's plot of annealed films on quartz substrate. (c) MZO1 and (d) MZO4.

The optical band gap of the MZO films was estimated by employing the Tauc model:

$$(\alpha h \nu) = A(h\nu - E_g)^{1/2} \quad \ldots (2)$$

where $\alpha$ is the absorption coefficient, $h\nu$ is the photon energy, $A$ is a constant and $E_g$ is the optical bandgap. The optical bandgap of the MZO films annealed at different temperatures was determined by extrapolation of the straight section to the energy axis of the plot of $(\alpha h\nu)^2$ versus photon energy $(h\nu)$ (Fig. 3c and d). The extrapolated bandgap values of MZO films showed systematic decrease with increasing annealing temperature. The lowering of the optical band gap value might be attributed to the reduced defect of the thin films with the augmenting annealing temperature. Moreover, MZO thin films exhibited higher optical band gap with ascending Mg concentration (Table 1). We attribute the band gap augmentation to the Burstein-Moss effect caused by an increase in free electron concentration due to the Mg doping.

## 4. Conclusions

We have studied the structural, topographical and optical properties of sol-gel derived MZO thin films deposited on quartz substrate after annealing at different temperatures. The investigation showed that the **lower boundary of the** grain size of all the films was increased with annealing temperature thus the crystallinity. The analysis also revealed that surface roughness of MZO films was increased upon annealing and this was responsible for the reduction in transmittance of the films.

**Acknowledgements**

We are grateful to Dr. S. Chattertjee from the department of Electronics and Communications Engineering, SMIT for his help with the spin coating. We would also like to thank Mr. N. Paitya and Mr. H. Pal from the department of Applied Electronics and Instrumentation, SMIT for their help with UV-Vis measurements.